\documentclass[reprint, amsmath, amssymb, aps, twocolumn]{revtex4}

\usepackage{graphicx}
\usepackage{dcolumn}
\usepackage[arrow, matrix,curve]{xy}
\usepackage{hyperref}
\usepackage{bbm}

\newcommand{\ovl}[1] {\overline{#1}}
\newcommand{\wt}[1] {\widetilde{#1}}
\newcommand{\wh}[1] {\widehat{#1}}
\newcommand{\cp} {\circ}

\newcommand{\SO} {\mathrm{SO}}

\newcommand{\quot} {\mathbf{q}}

\newcommand{\ot} {\otimes}
\newcommand{\vp} {\varphi}
\newcommand{\id} {\mathrm{id}}
\newcommand{\inn} {\mathrm{i}}
\newcommand{\pr} {\mathrm{p}}
\newcommand{\prr} {\mathrm{pr}}
\newcommand{\III} {\mathcal{S}}

\newcommand{\pap} {\pi_\AP}

\newcommand{\ID} {\mathcal{I}}

\newcommand{\gauss} {\sigma}

\newcommand{\p} {\widehat{p}}
\newcommand{\B} {\mathfrak{d}}
\newcommand{\oB} {\ovl{\B}}

\newcommand{\VDZ} {\mathfrak{e}_\AP}

\newcommand{\scal} {\mathrm{dil}}
\newcommand{\Scal} {\mathrm{Dil}}

\newcommand{\muRB} {\mu_{\mathrm{B}}}
\newcommand{\Gel} {\mathcal{G}}

\newcommand{\DIDE} {\mathfrak{D}}
\newcommand{\DIDEZ} {\mathfrak{d}_\AP}
\newcommand{\DIDEE} {\mathfrak{d}_0}

\newcommand{\E} {\mathbbm{1}}

\newcommand{\spann} {\mathrm{span}}

\newcommand{\CAPP} {C_{\mathrm{AP}}(\RR^3)}
\newcommand{\CAP} {C_{\mathrm{AP}}(\RR)}
\newcommand{\AP} {\mathrm{AP}}

\newcommand{\RR} {\mathbb{R}}
\newcommand{\RB} {\mathbb{R}_{\mathrm{Bohr}}}
\newcommand{\RBB} {\mathbb{R}^3_{\mathrm{Bohr}}}
\newcommand{\NN} {\mathbb{N}}

\newcommand{\CC} {\mathbb{C}}

\newcommand{\w} {\omega}

\newcommand{\he} {\hspace{1pt}}

\newcommand{\Hil} {\mathcal{H}}

\newcommand{\cA} {\mathfrak{a}}
\newcommand{\aA} {\mathfrak{A}}
\newcommand{\cC} {\mathfrak{C}}

\newcommand{\Spec} {\mathrm{Spec}}

\newcommand{\dd} {\mathrm{d}}
\newcommand{\ee} {\mathrm{e}}
\newcommand{\I} {\mathrm{i}}

\newcommand{\Borel} {\mathfrak{B}}

\newcommand{\diag}{\mathrm{diag}}
\newcommand{\Diff}{\mathrm{Diff}}

\begin{document}

\title{Uniqueness of the Representation in Homogeneous Isotropic LQC}

\author{Jonathan Engle${}^{1,2}$\footnote{jonathan.engle@fau.edu}, 
Maximilian Hanusch${}^1$\footnote{hanuschm@fau.edu}, 
Thomas Thiemann${}^2$\footnote{thomas.thiemann@gravity.fau.de}}
\affiliation{${}^1$ Department of Physics, Florida Atlantic University, 777 Glades Road, Boca Raton, FL 33431, USA\\
${}^2$ Institute for Quantum Gravity, Friedrich-Alexander University Erlangen-N{\"u}rnburg, Staudstra{\ss}e 7, 
91058 Erlangen, Germany}

\date{March 28, 2018}

\begin{abstract} 
We show that the standard representation of homogeneous isotropic loop quantum cosmology (LQC) is the GNS-representation that corresponds to the unique state on the reduced quantum holonomy-flux $^*$-algebra that is invariant under residual diffeomorphisms
---  \textit{both} when the standard algebra is used as well as when one uses the extended algebra proposed by Fleischhack. 
More precisely, we find that in both situations the GNS-Hilbert spaces coincide, and that in the Fleischhack case the additional algebra elements are just mapped to zero operators.
In order for the residual diffeomorphisms to have a well-defined action on the quantum algebra, 
we have let them act on the fiducial cell as well as on the dynamical variables, thereby recovering covariance. 
Consistency with Ashtekar and Campiglia in the Bianchi I case is also shown.
\end{abstract}

\maketitle

\section{Introduction}
\label{sdsdsdsd}

Diffeomorphism symmetry is the fundamental gauge symmetry which led Einstein to his unreasonably successful
theory of general relativity, and is the key symmetry which guides the quantization of this theory known as 
loop quantum gravity \cite{thiemann2007, rovelli2004, BackLA}.  In order to make contact between loop quantum gravity and observational predictions, as well as 
to test the classical limit of the theory in a simplified context, the framework of loop quantum cosmology was developed. There one starts with the homogeneous and (optionally) isotropic phase space of general relativity and quantizes using methods 
as close as possible to those used in loop quantum gravity.  When homogeneity is imposed, almost all diffeomorphism symmetry
is thereby automatically fixed, except for a three parameter family of \textit{residual diffeomorphisms}.  When
isotropy is additionally imposed, the residual diffeomorphisms are further reduced to the one parameter group of isotropic dilations.

In the quantization of the homogeneous models, a key technical subtlety is encountered: 
Because all dynamical fields become constant in space, 
the action integral, and hence the symplectic structure and Hamiltonian derived from it, diverges. As a consequence,
it is necessary to fix a finite cell over which to integrate in order to have a well-defined framework.  This choice of a cell artificially
breaks the residual diffeomorphism symmetry further.  In particular, only \textit{volume preserving} residual diffeomorphisms
preserve the resulting classical Poisson algebra, and hence only these preserve the resulting quantum algebra of basic operators.
%

In  the work of Ashtekar and Campiglia \cite{UniquOfKinLQC}, invariance under residual diffeomorphisms is used to select uniquely 
a representation of the reduced quantum holonomy-flux $^*$-algebra, in analogy to what is done in the full theory using 
the full diffeomorphism group \cite{LOST}. Because only volume preserving residual diffeomorphisms
have a well defined action on the reduced algebra, 
and such diffeomorphisms are non-trivial only in the non-isotropic case, 
their analysis is limited to a homogeneous, non-isotropic case -- specifically Bianch I. 

However, it is important to extend their uniqueness result to the isotropic case for two reasons (1.) the isotropic model is the
one that is actually being used to make contact with potential observations \cite{ABreview, AAN2013,  BCGM2014, BCT2011}, 
%
%
and (2.) thus far it is only in the isotropic 
case that the more general framework introduced by Fleischhack is available \cite{CSL}.
The Fleischhack framework improves upon the standard framework in the sense that it is more faithful to the quantization
procedure used in the full theory, in that the configuration algebra includes holonomies along \textit{all} analytic curves and not just those along straight ones.

In the prior work \cite{JM}, uniqueness was achieved for the isotropic case by focussing attention on selecting only 
the \textit{measure} on the quantum configuration space, and hence the inner product, using
invariance under residual diffeomorphisms, because there is no obstacle to defining the action of dilations
on the configuration space alone.  In this paper we propose a different solution to the problem which will allow for 
a stronger uniqueness result, namely uniqueness of the full representation of the quantum algebra, similar to what
is done by Ashtekar and Campiglia for the homogeneous case and by the authors of \cite{LOST} for the full theory. Specifically, we address the problem at its source: By allowing dilations to act on the fiducial cell, covariance is recovered, and an action of dilations which is well-defined on the quantum algebra is achieved. We show that there is 
only one state on both the standard algebra $\aA_s$ and on the Fleischhack algebra $\aA$ that is invariant under this action of dilations. We find that in the standard case, the corresponding GNS-representation is unitarily equivalent to the standard representation $\varrho_s$ used in LQC up until now. In the Fleischhack case, we have a canonical embedding $\iota\colon \aA_s\rightarrow \aA$, 
where $\aA= \iota(\aA_s)\oplus\ID_0$ holds for $\ID_0$ the ideal generated by the additional elements coming from parallel transports along non-straight curves. Then, for $\varrho$ the GNS-representation that corresponds to the invariant state on $\aA$, (up to unitary equivalence) we have 
\begin{align*}
	\varrho\cp \iota=\varrho_s\qquad\text{and}\qquad\varrho|_{\ID_0}=0,
\end{align*} 
i.e., $\varrho$ is actually just given by its restriction to the standard holonomy flux $^*$- algebra of homogeneous isotropic LQC and equals the standard representation thereon.
%
%

\section{SYMMETRY REDUCTION, CANONICAL ANALYSIS, AND DILATIONS}
\label{sec:classical}

In this section, we derive the symmetry reduction of the Holst action and perform the Legendre transform
in order to obtain the form of the canonical variables in the spatially flat, homogeneous isotropic case.  As far as we are aware,
this is the first time the reduced canonical framework has been derived in this way. This derivation serves to fix conventions
as well as to carefully derive from first principles the various factors involved.  

Loop quantum gravity is based on the Holst action of gravity \cite{holst1995}, in which the basic variables 
are an $\SO(1,3)$ connection $\omega_\alpha^{IJ}$ and a co-tetrad $e^I_\alpha$. Here,  
$I,J = 0, \dots, 3$ denote internal indices which are raised and lowered using $\eta_{IJ}:= \diag(-1,1,1,1)$.
To impose homogeneity and isotropy, one fixes a global foliation of space-time $\mathcal{M}$ into spatial 3-slices.
One then chooses one of the possible isometry groups of homogeneity and isotropy in three dimensions, 
and fixes some action of this group on $\mathcal{M}$ via diffeomorphisms preserving each slice.
In this paper we choose our symmetry group to be the one so far suggested by observation --- the Euclidean group 
$\mathcal{E}= \RR^3 \rtimes \SO(3)$ (i.e., the $k=0$, spatially flat case)  
with multiplication $(u,r)\cdot(u',r') = (u+ru', rr')$, so that the 3-slices of $\mathcal{M}$ are diffeomorphic to $\RR^3$. 
Moreover, we introduce a global chart $(x^\mathbf{\alpha}) = (x^\mathbf{0}, x^\mathbf{a})$ such that $x^\mathbf{0}$ is constant on each spatial slice, and the action $\mathcal{E} \ni (u,r) \mapsto \Phi_{(u,r)} \in \Diff(\mathcal{M})$  
takes the standard form
\begin{align*}  
	\Phi_{(u,r)}((x^\mathbf{0}, x^\mathbf{a}))=(x^\mathbf{0}, u^\mathbf{a}+r^\mathbf{a}{}_\mathbf{b}\hspace{0.5pt} x^\mathbf{b}).
\end{align*}
If one requires invariance of the basic variables $(\omega_\alpha^{IJ}, e_\alpha^I)$ under an action of $\mathcal{E}$ via diffeomorphisms alone \footnote{The basic fields in this section are naturally understood to be associated to an $\mathrm{SO}(1,3)$ principal fiber bundle $P$ over $\mathcal{M}$ which we assume to be trivial, and for which we fix a global trivialization $P = \mathcal{M} \times \mathrm{SO}(1,3)$ .  By ``diffeomorphisms alone'', we mean lifts of elements of $\mathrm{Diff}(\mathcal{M})$ to $P$ which act only on the $\mathcal{M}$-factor.}, the resulting solution space is empty; as a consequence, in order to have non-trivial invariant basic variables, it is necessary for $\mathcal{E}$ to act via gauge rotations as well.  
Specifically, for each $r \in \SO(3)$ define
\begin{align*}
\Lambda(r)^I{}_J :=  \left( \begin{array}{cc} 1 & \vec{0}^T \\ \vec{0} & r^i{}_j \end{array} \right) \in \SO(1,3),
\end{align*}
and let each element $(u,r) \in \mathcal{E}$ act on the basic variables via \cite{engle2007, BojoKa}
$\Lambda(r) \circ \Phi_{(u,r)}$. Here, gauge transformations and diffeomorphisms are defined to act on the basic variables 
in such a way that their action is a left action. 
Let $\III$ denote the space of all pairs $(\omega,e)$ invariant under this action. They are 
exactly of the form
\begin{align}
\label{invform}
\begin{aligned}
\omega_{\mathbf{a}}^{ij} = w\hspace{1pt} \epsilon_{\mathbf{a}ij}, 
\quad \omega_\mathbf{a}^{0i} = - \frac{c}{\gamma}\hspace{1pt}\delta^i_\mathbf{a},
\quad \omega_\mathbf{0}^{ij} = 0, \quad \omega^{0i}_\mathbf{0} = 0, \\
e^0_\mathbf{a} = 0, \quad  e^0_\mathbf{0} = N, \quad  e^i_\mathbf{0} = 0, \quad
e^i_\mathbf{a} = v \hspace{1pt} \delta^i_\mathbf{a},\quad\:\:
\end{aligned}
\end{align}
for some smooth real valued functions $w, c , v, N$, depending on $x^\mathbf{0}$ only. 
At this point, let us additionally assume $N > 0$ (the time orientation part of the time gauge).
%
%

The (dynamically oriented) Holst action is given by 
\begin{align*}
S_H = 
\frac{1}{4k}\! \underset{(\mathcal{M}, [e])}{\int} \!\!\! \left( \epsilon_{IJKL}\: e^I  \wedge  e^J \wedge  \Omega^{KL} 
\! - \! \frac{2}{\gamma}\: e^I  \wedge  e^J \wedge  \Omega_{IJ} \right) 
\end{align*}
where $k:= 8\pi G$, $\gamma$ is the Barbero-Immirzi parameter \cite{barbero1995, immirzi1995}, $\Omega^{IJ}$ 
is the curvature of $\omega^{IJ}$, the four dimensional alternating tensor is defined such that 
%
%
$\epsilon^{0123} = 1$ so that
$\epsilon_{0123}=-1$, and 
where $(\mathcal{M}, [e])$ denotes $\mathcal{M}$ with the orientation defined by the co-tetrad $e$. 
The Legendre transform of this action yields the canonical variables 
$A^i_a = \Gamma^i_a + \gamma K^i_a$, the Ashtekar connection, and 
$E^a_i = |\det e^j_b|\hspace{1pt} e^a_i$. Here, $\Gamma_a^i$ denotes the spin connection determined by $e_a^i$,
and $K^i_a := \omega^{i0}_a$. 
Note in particular that the definition of the orientation of the action integral used here
makes it independent of any background orientation. 
More importantly, this choice of orientation is necessary to ensure that the variable conjugate to $A^i_a$ is
$E^a_i$ defined with absolute value around the determinant of the triad, which ensures equivalence of the framework with the generalized ADM framework \cite{thiemann2007, BackLA}. 

Let $\mathcal{L}_{H}$ denote
the integrand of this action, i.e., the Holst Lagrangian density. 
Evaluated on $\III$,  this Lagrangian density reduces to
\begin{align*}
\mathcal{L}_{H} = \frac{3 |v|v }{k\gamma}\:\frac{\mathrm{d}}{\mathrm{d}x^{\mathbf{0}}}(c-w) 
-\frac{3N|v|}{k\gamma^2}(c^2 - \gamma^2 w^2 - 2cw).
\end{align*}
As $\mathcal{L}_{H}$ is constant in space, the corresponding Lagrangian $L_{H}$ must be defined by integrating over a finite cell 
$\mathcal{C}$ in space.  Let $V_0[\mathcal{C}]$ denote the unsigned coordinate volume of the cell $\mathcal{C}$ in the fixed
coordinates $x^{\mathbf{a}}$, so that 
\begin{align*}
L_{H} \! = \! \frac{3 V_0[\mathcal{C}] |v|v }{k\gamma}\frac{\mathrm{d}}{\mathrm{d}x^{\mathbf{0}}}(c\!-\!w) 
-\frac{3N V_0[\mathcal{C}] |v|}{k\gamma^2}(c^2\! - \! \gamma^2 w^2 \! - \! 2cw).
\end{align*}
The symplectic term in this Lagrangian tells us that the dynamical variable is $c-w$, 
with canonically conjugate momentum
\begin{align}
\label{pdef}
p:= \frac{3 V_0[\mathcal{C}] |v|v }{k\gamma},
\end{align}
which motivates a change of variables 
from $(c,w)$ to 
\begin{align*}
c_{\pm} := c \pm w;
\end{align*}
yielding 
\begin{align*}
 L_{H} = p\hspace{1pt}\dot{c}_- \hspace{1pt}
+\: &   N \sqrt{\frac{3 V_0[\mathcal{C}]}{16 k\gamma^3}} \,\,  |p|^{1/2}
\big( (1+\gamma^2) \hspace{1pt} c_+^2  \\
& \qquad \quad - (3-\gamma^2)\hspace{1pt} c_-^2 - 2(1+\gamma^2)\hspace{1pt} c_+ c_-\big).
\end{align*}
Here, $N$ and $c_+$ have no time derivatives in the Lagrangian, and so are Lagrange multipliers. 
The constraint obtained by varying $c_+$ is
\begin{align}
\label{secclass}
c_+- c_- = 0.  
\end{align}
In the Dirac constraint analysis of this system \cite{dirac1964}, the above constraint turns out to be second class, 
for reasons similar to that used in \cite{holst1995}.
%
%
Specifically, note that the canonical momentum $\Pi_{c_+}$ conjugate to $c_+$ is trivially constrained to be zero
$\Pi_{c_+} \approx 0$. The canonical Poisson brackets then yield
$\{c_+-c_-, \Pi_{c_+}\} = 1$, so that both $c_+-c_- \approx 0$ and $\Pi_{c_+} \approx 0$ are second class.
Thus both are substituted into the Lagrangian prior to quantization.
Doing this, and writing the Lagrangian in terms of the original variables, we have 
\begin{align*}
 L_{H} = p\hspace{1pt}\dot{c} 
- N \sqrt{\frac{3 V_0[\mathcal{C}]}{k\gamma^3}} \,  |p|^{1/2} c^2, 
\end{align*}
from which one can read off the conjugate variables $c$ and $p$, as well as the usual Hamiltonian constraint
of isotropic LQC \cite{as2011a}.
The Ashtekar connection $A^i_a$, and densitized triad $E^a_i$, when evaluated on $\III$, in terms of $c$ and $p$,
are of the form
\footnote{These hold only in the fixed coordinate system $x^{\mathbf{a}}$. 
In order to have expressions valid in any coordinate system,
it is sufficient to recast them in a form covariant under coordinate transformation.  
This can be done by introducing a fiducial triad and co-triad via 
$\mathring{e}^a_i := \left(\frac{\partial}{\partial x^{\mathbf{i}}}\right)^a$ and
$\mathring{e}^i_a := (dx^{\mathbf{i}})_a$.
$V_0[\mathcal{C}]$ is then the volume of $\mathcal{C}$ with respect to $\mathring{q}_{ab}:=\mathring{e}^i_a\mathring{e}_{bi}$,
and the expressions for $A^i_a$ and $E^a_i$ take the covariant forms
$A^i_a = c \hspace{1pt}\mathring{e}^i_a$ and 
$E^a_i :=  \frac{k\gamma}{3 V_0[\mathcal{C}]} \hspace{1pt} p \hspace{1pt} \det(\mathring{e}^j_b)\hspace{1pt} \mathring{e}^a_i$.
}
\begin{align}
\label{AErelation}
 A^i_\mathbf{a} = c \hspace{1pt}\delta^i_\mathbf{a}, \qquad
E^\mathbf{a}_i = \frac{k\gamma}{3 V_0[\mathcal{C}]} \hspace{1pt} p \hspace{1pt} \delta^\mathbf{a}_i .
\end{align}

\subsection*{THE ACTION OF DILATIONS}

Since the theory is restricted to a cell, and dilations do not preserve this cell, dilations are in fact not defined in the theory, \textit{unless we let the dilations act on the cell as well}.  
%
%
This is the key new ingredient which will let dilations have a well-defined action on the quantum algebra.
We begin by deriving the action of a dilation on $c$ and $v$. For this, we let $\Phi_\lambda$ denote the dilation which expands space from the origin by a factor of $\lambda\in \RR_{>0}$, as measured in the 
fixed coordinate system $x^{\mathbf{a}}$. 
Then, letting $\Phi_\lambda$ act on the cell $\mathcal{C}$ transforms
$V_0[\mathcal{C}]$ to $|\lambda|^3 V_0[\mathcal{C}]$. Moreover, the $1$-forms $\w$ and $e$ transform via push forward by $\Phi_\lambda$ (pull-back by $\Phi_{\lambda^{-1}}$) \footnote{Since $\Phi_\lambda$ acts on curves via $\gamma\mapsto\Phi_\lambda\cp \gamma$, it acts on vector-fields via push-forward by $\Phi_\lambda$ (recall the definition of a tangent vector as a equivalence class of curves). 
Since $\Phi_\lambda$ also acts on scalar fields $\phi$ via push-forward $[\Phi_\lambda]_*(\phi)= \phi\cp\Phi_{\lambda^{-1}}$, and since the natural pairing of a vector-field with a $1$-form is a scalar field, $1$-forms must transform via push-forward by $\Phi_\lambda$ as well. Moreover, it is with this action that standard constructions such as form integrals are diffeomorphism covariant.}, whereby
\begin{align*}
	[\Phi_\lambda]_*\colon \III&\rightarrow \III\\
	 (\w,e)&\mapsto ([\Phi_\lambda]_*(\w),[\Phi_\lambda]_*(e))
\end{align*}
is easily seen to map $\III$ to $\III$ (recall that the quantities in \eqref{invform} only depend on $x^{\mathbf{0}}$). More concretely, we have that  
 $(v,c)$ transforms to $(\lambda^{-1} v, \lambda^{-1} c)$ \footnote{To see this 
quickly, let $\ell$ denote the straight spatial curve $x^{\mathbf{a}}(t) := t\hspace{1pt} \ell^{\mathbf{a}}$, $t\in (0,1)$, so that 
$\int_\ell \omega^{0i} = - (c/\gamma)\hspace{1pt} \delta^i_\mathbf{a} \ell^\mathbf{a}$ and 
$\int_\ell e^i = v \hspace{1pt}\delta^i_\mathbf{a} \ell^\mathbf{a}$.  
The diffeomorphism covariance of the form integral implies that both of these integrals remain unchanged when the dilation $\Phi_\lambda$ acts on $\omega^{0i}$, $e^i$, and $\ell$ simultaneously.
But under this dilation $x^{\mathbf{a}}$ transforms to $\lambda\hspace{1pt} x^{\mathbf{a}}$, so that  
$c$ and $v$ transform as claimed. 
}, so that, by equation (\ref{pdef}), $p$ transforms
to $\lambda \hspace{1pt} p$. That is, under the dilation $\Phi_{\lambda}$, we have  
\begin{align}
\label{transfff}
	(p,c) \mapsto (\lambda\hspace{1pt} p,\lambda^{-1}c).
\end{align}

Now, 
the elementary variables of the quantum theory are  
$p$ and (a suitable dense subalgebra of) the complex $C^*$-algebra $\cC$ generated by parallel transports of the Ashtekar connection, 
as determined by $c$ via (\ref{AErelation}), along a certain class of curves. $\cC$ is referred to as the configuration algebra.
Depending on whether the curves are restricted to be straight (standard LQC \cite{MathStrucLQC}) or include all embedded analytic curves (Fleischhack approach \cite{CSL}), we have
\begin{equation}
  \label{eq:Ksiiii}
  \cC = 
  \begin{cases} 
    \hspace{38.8pt} \CAP &\:\:\mbox{ standard LQC}\\ 
    C_0(\RR)\oplus \CAP &\:\:\mbox{ Fleischhack approach}.
  \end{cases}
\end{equation}
Here, $C_0(\RR)$ denotes the set of all continuous functions on $\RR$ that vanish at infinity, and $\CAP$ the space of  almost periodic functions on $\RR$, i.e.,  the $C^*$-subalgebra of the bounded functions on $\RR$ that is generated by the characters
 \begin{align*}
 	\chi_ \lambda\colon \RR\mapsto \CC,\quad t\mapsto\ee^{\I\lambda t}\qquad\forall\: \lambda\in \RR.
 \end{align*} 
The action of dilations on these elementary variables is, from \eqref{transfff},
\begin{align}
\label{fdifdsioiu}
	\textstyle(p,\vp)\mapsto (\lambda\cdot p, \vp_\lambda)\quad\:\:\text{for}\quad\:\: \vp_\lambda\colon t\mapsto \vp(\lambda^{-1}\cdot t).
\end{align}
\section{QUANTIZATION}
\label{quanti}
The elementary variables of homogeneous isotropic LQC are the  
elements of a suitable dense subalgebra $\DIDE$ of the configuration algebra \eqref{eq:Ksiiii}
 together with the momentum $p$. More precisely, the classical Poisson algebra is the  complex vector space $\CC\cdot p \times \DIDE$ with Lie bracket 
\begin{align}
\label{dsopspod}
\{(z p,\vp),(z'p,\vp')\}=\{0,z'\dot\vp-z\dot\vp'\},
\end{align}
where $\dot\varphi$ denotes the derivative of $\vp$ with respect to its argument. 
In the standard case, we choose $\DIDE$ to be the largest subalgebra  of $\cC$ for which \eqref{dsopspod} makes sense, namely, 
\begin{align*}
	\DIDEZ:=\{\vp\in \CAP\cap C^\infty(\RR)\:|\: \vp^{(n)}\in \CAP\:\:\forall\in \NN\}.
\end{align*}
In the Fleischhack case, we define $\DIDE:=\DIDEE\oplus \DIDEZ$ for
\begin{align*}
	\DIDEE:=\{\vp\in C_0(\RR)\cap C^\infty(\RR)\:|\: \vp^{(n)}\in C_0(\RR)\:\:\forall\in \NN\}.
\end{align*}
In both cases, $\DIDE$ is dense in $\cC$. This is clear in the standard case, because 
\begin{align}
\label{uoioioi}
	\VDZ:=\spann_\CC(\{\chi_\lambda\}_{\lambda\in \RR})\subseteq \DIDEZ
\end{align}
is dense in $\CAP$ by definition. In the Fleischhack case, we define $C_0(\RR)\ni\gauss:=\gauss_0$, with
\begin{align}
\label{gaussss}
\gauss_\epsilon\colon \RR\rightarrow \RR,\quad t\mapsto \ee^{-(\epsilon+t)^2}
\end{align}
for $\epsilon\in \RR$, 
and observe that the products $\{\chi_\lambda\cdot \gauss\}_{\lambda\in \RR}$ separate the points in $\RR$; thus, generate a dense $^*$-subalgebra of $C_0(\RR)$ by the Stone-Weierstrass theorem as $\gauss$ vanishes nowhere.

\subsection*{THE QUANTUM ALGEBRA} 
We now define the reduced quantum holonomy-flux $^*$-algebra $\aA$ over the basic variables $\p$ and $\wh{\vp}$ for $\varphi\in \DIDE$, that, setting $\hbar=1$, we want to obey the commutation relations 
\begin{align}
\label{basiccomm}
\textstyle[\p, \widehat{\varphi}] = - \I \cdot \widehat{\dot\varphi}\qquad\forall\:\varphi\in \DIDE;
\end{align}
the quantum analogue to \eqref{dsopspod} for $p\equiv(p,0)$ and $\vp\equiv(0,\vp)$. In addition to that, we want to implement
\begin{align*}
	\wh{\vp}'\wh{\vp}=\wh{\vp'\vp}\qquad\forall\: \vp,\vp'\in \DIDE.
\end{align*} 
For this, let $\cA$ denote the (mixed) complex tensor algebra over the complex vector spaces $\CC\cdot p$ and $\DIDE$. We define the involution $*\colon \cA\rightarrow\cA$ by antilinear extension of
\begin{align*}
	(o_1\ot{\dots}\ot o_n)^*:=o^*_n\ot{\dots}\ot o^*_1 
\end{align*}
with $\vp^*:=\ovl{\vp}$ for each $\vp\in \DIDE$, $(z p)^*:=\ovl{z} p$ for each $z\in \CC$, as well as $\ovl{1}_\cA:=1_\cA$. Then, the  
reduced quantum holonomy-flux $^*$-algebra $\aA$ is defined to be the quotient of 
$\cA$ by the both sided $^*$-ideal $\ID\subseteq \cA$ that is generated by the elements
\begin{align*}
	 p\ot\varphi-\varphi&\ot p+\I\cdot\dot{\varphi}\\[2pt]
	 	\varphi_1\ot\varphi_2 &- \varphi_1\varphi_2\\[2pt]
	 	\chi_0 &- 1_\cA 
\end{align*}
for all $\varphi,\varphi_1,\varphi_2\in \DIDE$. We then define $\p:=[ p]$, $\E:=[ \chi_0]$, as well as $\wh{\vp}:=[ \vp]$ for each $\vp\in \DIDE$.   
\vspace{6pt}

\noindent
Finally, we impose the action of residual diffeomorphisms 
on $\aA$ as follows. For each $\lambda\in \RR_{> 0}$, we define  
$\scal_\lambda\colon \cA\rightarrow \cA$ by linear extension of 
\begin{align*}
	\scal_\lambda\colon o_1\ot{\dots}\ot o_n\mapsto \scal_\lambda(o_1)\ot{\dots}\ot \scal_\lambda(o_n),
\end{align*}
with $\scal_\lambda(z p):=\lambda\cdot z p$ for each $z\in \CC$, 
\begin{align*}  
	\scal_\lambda(\varphi)\colon t\mapsto \vp(\lambda^{-1}\cdot t)\quad \forall\:\varphi\in \DIDE,
\end{align*} 
as well as $\scal_\lambda(1_\cA):=1_\cA$, cf.\ \eqref{fdifdsioiu}. 
Since $\scal_\lambda$ preserves the algebraic structure of $\cA$ as well as the ideal $\ID$, it carries over to a well-defined homomorphism $\Scal_\lambda\colon \aA\rightarrow \aA$. 
\subsection*{STATES: THE GNS CONSTRUCTION}
Given a unital $^*$-algebra $\aA$, a {\bf state} (on $\aA$) is a linear functional $\w\colon \aA\rightarrow \CC$  with $\w(a^*a)\geq 0$ and $\w(\E)=1$. Here, we automatically have the Cauchy-Schwarz inequality
\begin{align}
\label{CASC}
	|\w(a^*b)|^2&\leq \w(a^*a)\cdot \w(b^*b)
\end{align}
for each $a,b\in \aA$. The corresponding Gel'fand (left) ideal is defined by $\ID_\w:=\{a\in \aA\:|\: \w(a^*a)=0\}$ and provides us with the pre-Hilbert space $(\Hil_\w,\langle\cdot,\cdot\rangle_\w)$, for $\Hil_\w$ the quotient of $\aA$ by $\ID_\w$, and 
\begin{align*}
	\langle \psi_a,\psi_b\rangle_\w:=\w(a^*b)\qquad\forall\:\psi_a,\psi_b\in \Hil
\end{align*}
 ($\psi_a$ denotes the class of $a\in \aA$ in $\Hil_\w$). In addition to that, we have the $^*$-representation 
	$\varrho_\w\colon \aA\rightarrow \Borel(\Hil_\w)$, defined by $\varrho_\w(a)(\psi_b):=\psi_{ab}$. Here, and in the following, $\Borel(\Hil)$ denotes the space of adjointable operators on the pre-Hilbert space $\Hil$. 
	
	An important feature of the GNS-construction is that $\psi_\E$ is  cyclic, i.e., we have $\varrho_\w(\aA)(\psi_\E)=\Hil$. Moreover, if $(\varrho,\Hil,\langle\cdot,\cdot\rangle)$ is a further $^*$-representation of $\aA$ on a pre-Hilbert space $\Hil$, with 
\begin{align*}
	\w(a)= \langle\psi,\varrho(a)(\psi)\rangle\qquad \forall\: a\in \aA 
\end{align*}
for a cyclic vector $\psi\in \Hil$, then $(\varrho_\w,\Hil_\w,\langle\cdot,\cdot\rangle_\w)$ is unitarily equivalent to
$(\varrho,\Hil,\langle\cdot,\cdot\rangle)$ via the intertwiner
\begin{align*}
		 \Hil_\w\rightarrow \Hil,\quad \psi_a\mapsto \varrho(a)(\psi).
\end{align*} 

\section{UNIQUENESS OF THE INVARIANT STATE}
\label{jndshjsd}
In the following, let $\aA$ denote the quantum holonomy-flux $^*$-algebra  defined in the first subsection of Sect.\ \ref{quanti}. We first show a continuity property of states on $\aA$, and then prove uniqueness and existence of the dilation-invariant state on $\aA$. Then, we show that the GNS-representation of this state is unitarily equivalent to the standard representation of homogeneous isotropic LQC in both the standard and the Fleischhack case. More precisely, we find that the GNS-Hilbert space is the same in both situations, and that in the Fleischhack case, the additional algebra elements are just mapped to zero operators. 

\subsection*{CONTINUITY}
Let $\w\colon \aA\rightarrow \CC$ be a state, and let $\B=\DIDEZ$ in the standard case, as well as $\B=\DIDEE$ or $\B=\DIDEZ$ in the Fleischhack one. Moreover, denote by $\oB$ the closure of $\B$ in $\cC$, i.e.,
\begin{equation*}
  \oB = 
  \begin{cases} 
    \CAP &\mbox{if }\: \B=\DIDEZ\\ 
    \hspace{7.4pt}C_0(\RR)  & \mbox{if }\: \B=\DIDEE.
  \end{cases}
\end{equation*}
We claim that the linear functional
\begin{align}
\label{fdgdfgg}
	\w_{\B}\colon \B \rightarrow \CC,\quad \vp\mapsto \w(\wh{\vp})
\end{align}  
is continuous w.r.t.\ $\|\cdot\|_\infty$, i.e., that it extends to a continuous linear functional $\w_{\oB}\colon \oB\rightarrow \CC$. 
To see this, first observe that the smooth function \footnote{As figured out in \cite{CSL2}, in the previous versions of this article  it had been overseen that, instead of $\nu\cp \vp\in C_0(\RR)$, we have $\nu\cp \vp\in \spann_{\CC}(\chi_0) + C_0(\RR)$ for $\vp\in C_0(\RR)$ and $\nu\colon (-1,1)\ni t\mapsto \sqrt{1+t}$. 
In the current version, we have fixed this issue by adding ``$-1$'' in the definition of $\nu$. 
We then have $\sqrt{1+\phi} \in \DIDE$ (instead of $\sqrt{1+\phi} \in \B$, as wrongly stated in the previous versions of this article) -- This, however, makes no difference for our conclusion that $1+\w(\vp_\nu)\geq 0$ holds, as $\w$ is defined on full $\DIDE$.}
\begin{align*}
	\nu\colon (-1,1)\ni t\mapsto \sqrt{1+t}-1
\end{align*}
(and each of its derivatives) can be represented by a power series.   
Thus, by completeness of $\oB$, for
$\vp\in \oB$ real valued with $\|\vp\|_\infty<1$, we have $\nu\cp \vp\in \oB$ 
as well as 
\begin{align*}
	\nu^{(n)}\cp \vp\in \spann_{\CC}(\chi_0) + \oB\qquad\forall\: n\geq 1.
\end{align*}
Then, for $\vp\in \B$, the chain rule gives 
\begin{align*} 
	\textstyle\partial_t (\nu\cp \vp)=(\dot\nu\cp\vp)\cdot \dot{\vp}\in \ovl{\B}.
\end{align*}
Now, the right hand side is differentiable; and, applying the same arguments inductively, we find that $(\nu\cp\vp)^{(n)}\in \ovl{\B}$ holds for each $n\in \NN$, hence
\begin{align*}
	\nu\cp \vp=\sqrt{1+\vp}-\chi_0\in \B.
\end{align*}
We thus have $\vp_\nu:= \sqrt{1+\vp} \in \DIDE$ with
\begin{align*}
	1+\w_\B(\vp)&= \w(\E+\wh{\vp})=\w(\wh{\vp}_\nu^*\wh{\vp}_\nu)\geq 0.
\end{align*}
Then, for each non-zero and real-valued $\vp\in \B$, we have
\begin{align*}
	\textstyle 1\pm \frac{\epsilon}{\|\vp\|_\infty}\cdot \w_\B(\vp)\geq 0\quad&\textstyle\Longrightarrow\quad \pm\phantom{|}\w_\B(\vp)\phantom{|}\geq -\textstyle\frac{1}{\epsilon}\cdot \|\vp\|_\infty\\
	&\textstyle\Longrightarrow\quad \phantom{\pm}|\w_\B(\vp)|\leq \textstyle\phantom{-}\frac{1}{\epsilon}\cdot \|\vp\|_\infty
\end{align*}
for each $0<\epsilon<1$.
This shows $|\w_\B(\vp)|\leq \|\vp\|_\infty$ for each real-valued $\vp\in \B$. Then, for $\vp\in\B$ arbitrary, we conclude from $\|\Re(\vp)\|_\infty,\|\Im(\vp)\|_\infty\leq \|\vp\|_\infty$ that
\begin{align*}
	|\w_\B(\vp)|\leq |\w_\B(\Re(\vp))| + |\w_\B(\Im(\vp))|\leq 2\cdot\|\vp\|_\infty.
\end{align*}
\subsection*{UNIQUENESS}
Suppose that $\w$ is invariant under dilations. Then, 
\begin{align*}
	\w(\p^*\p)=\w(\p^2)=\Scal_\lambda(\w)(\p^2)=\lambda^2\cdot \w(\p^*\p)
\end{align*}
for $0<\lambda\neq 1$, implies $\w(\p^*\p)=0$; hence, $\p\in \ID_\w$. Thus, we have
\begin{align}
\label{odpdpdd}
	\w|_{\aA\cdot \p}=0=\w|_{\p\cdot \aA}
\end{align} 
by \eqref{CASC}, as $\p^*=\p$. 
This implies that $\w$ is uniquely determined by its values on $\{\wh{\vp}\:|\: \vp\in \DIDE\}$, because, applying \eqref{basiccomm} successively, we see that each element in $\aA$ can be written as a sum of elements of $\{\wh{\vp}\:|\: \vp\in \DIDE\}$ and $\aA\cdot \p$. In addition to that, \eqref{odpdpdd} shows 
	$\w(a\cdot \p)=0=\w(\p\cdot a)$ for each $a\in \aA$; 
hence, $\w_\B\cp \partial_t=0$ as 
\begin{align*}
	-\I\cdot \w(\hspace{1pt}\wh{\dot\vp}\hspace{1pt})=\w([\p,\wh{\vp}])=0\quad\forall\:\vp\in \DIDE.
\end{align*} 
In particular, we have 
\begin{align}
\label{gfhghg}
	\textstyle\w_{\DIDEZ}(\chi_\lambda)=-\frac{\I}{\lambda}\cdot\big(\w_{\DIDEZ}\cp \partial_t\big)(\chi_\lambda)=0\quad\forall\:\lambda\neq 0,
\end{align}
as well as $\w_{\DIDEZ}(\chi_0)=\w(\E)=1$. By denseness of $\VDZ$ in $\DIDEZ$ and by continuity of $\w_{\DIDEZ}$, this completely determines $\w_{\oB_\AP}$. Thus, uniqueness is clear in the standard case.

 Furthermore, in the Fleischhack case, we must have $\w_{\oB_0}=0$, so that it reduces to the standard case.
 To see this, let us first observe that $\w_{\oB_0}$ is positive, because for $C_0(\RR)\ni\vp\geq 0$, we have $\sqrt{\vp}\in C_0(\RR)$; hence,
\begin{align*}
	\w_{\oB_0}(\vp)=\textstyle\lim_n \w_{\oB_0}(\ovl{\vp}_n \vp_n)=\lim_n \w(\wh{\vp}_n^* \wh{\vp}_n)\geq 0
\end{align*}
for $\{\vp_n\}_{n\in \NN}\subseteq \B_0$ a sequence with $\lim_n\vp_n=\sqrt{\vp}$. Thus,
\begin{align*}
	\textstyle\w_{\oB_0}\colon \vp\mapsto \int \vp\: \dd\mu
\end{align*}
holds for some unique finite Radon measure $\mu$ on $\RR$, by the Riesz-Markov theorem. Then, dilation-invariance of $\w$ implies 
$\mu=s\cdot \mu_\delta$ for some $s\geq 0$, with
\begin{equation*}
  \mu_\delta(A) := 
  \begin{cases} 
    1 &\mbox{if } 0\in A\\ 
    0  & \mbox{if } 0\notin A
  \end{cases}
\end{equation*}
for each $A\in \Borel(\RR)$, cf.\ \cite{JM}. Then,
\begin{align*}
	0=(\w_{\DIDEE}\cp\partial_t)(\gauss_\epsilon)=\w_{\DIDEE}(\dot\gauss_\epsilon)=s\cdot \dot\gauss_\epsilon(0)=-2s\epsilon\cdot \gauss(\epsilon)
\end{align*}
shows $s=0$; hence, the claim. Here, $\gauss_\epsilon$ is defined by \eqref{gaussss}.

\subsection*{THE STANDARD FUNCTIONAL}
For existence of the dilation-invariant state, we will need the linear functional
\begin{align}
\label{hggffhjhjkj}
	\textstyle L\colon \cC\rightarrow \CC,\quad \vp\mapsto \lim_n \frac{1}{2n}\int_{-n}^n \vp(t)\:\dd t,
\end{align}
whose well-definedness follows easily from
\begin{align}
\label{hjhjgfhj}
	L(\chi_0)=1\qquad\text{and}\qquad L(\chi_\lambda)=0\quad\forall\:\lambda\neq 0,
\end{align}
from denseness of $\VDZ$ in $\CAP$, as well as from the fact that $L|_{C_0(\RR)}=0$ holds in the Fleischhack case. 
$L$ is furthermore directly seen to be continuous. From this, and the fact that 
$\scal_\lambda$ is an isometry for each $\lambda> 0$, we have
\begin{align}
\label{ssddssdsaaa}
	L\cp\scal_\lambda = L\qquad\forall\:\lambda\in \RR_{> 0}
\end{align}
as this clearly holds on $\VDZ$. 
Moreover, it is clear that $\langle\vp_1,\vp_2 \rangle=\ovl{\langle\vp_2,\vp_1 \rangle}$ holds for
\begin{align}
\label{fddfgfsss}
	\langle\vp_1,\vp_2 \rangle:= L(\ovl{\vp}_1\vp_2)\qquad \forall\: \vp_1,\vp_2\in \cC;
\end{align}
and by continuity of $L$, we have $\langle\vp,\vp \rangle>0$ for each $0\neq\vp\in \CAP$, cf.\ Appendix A.  
Thus, $\Hil:= \DIDEZ$ is a pre-Hilbert space w.r.t.\ $\langle\cdot,\cdot\rangle$. 

\subsection*{EXISTENCE}
We now show existence of the dilation-invariant state for both the standard and the Fleischhack case simultaneously.  
For this, we let  $\pap\colon \cC\rightarrow \CAP$ denote the identity on $\CAP$ in the standard case, as well as the projection onto $\CAP$ in the Fleischhack one. We define the operators
\begin{align*}
	\wt{\vp}\colon \Hil\rightarrow\Hil,&\quad \vp'\mapsto \pap(\vp)\cdot \vp'\\
	\wt{z p}\colon\Hil\rightarrow\Hil,&\quad \vp'\mapsto -z\I\cdot \dot\vp'
\end{align*}
for each $\vp\in \DIDE$ and $z\in \CC$, as well as $\wt{z1}_\cA:=z\cdot \id_\Hil$ for each each $z\in \CC$. Obviously, then 
\begin{align}
\label{posdopsdpo}
	[\wt{p},\wt{\vp}]=-\I\cdot\wt{\dot\vp}\qquad\forall\:\vp\in \DIDE
\end{align}
holds. In addition to that, we have
\begin{align*}
	\langle\hspace{1pt} \wt{\ovl{\vp}}(\cdot),\cdot\rangle &= \langle \cdot,\wt{\vp}(\cdot)\rangle\qquad \forall\: \vp\in \DIDE\\
	\langle\wt{p}(\cdot),\cdot\rangle &= \langle \cdot,\wt{p}(\cdot)\rangle;
\end{align*}
whereby the first line is obvious, and the second one is due to partial integration. 
 From this, it follows inductively that
$\Omega\colon  \cA\rightarrow \CC$ defined by linear extension of 
\begin{align}
\label{pdisisad}
\Omega(o_1\ot{\dots}\ot o_n):=\langle \chi_0,(\wt{o}_1\cp{\dots}\cp \wt{o}_n)(\chi_0)\rangle
\end{align}
is positive. Moreover, it is straightforward from \eqref{posdopsdpo} 
that $\Omega$ is zero on $\ID$; thus, defines a positive linear functional $\w$ on $\aA$, for which obviously $\w(\p^*\p)=0=\w(\p)$ holds. In addition to that, it is immediate from \eqref{ssddssdsaaa}, the definition of $\wt{p}$, and the chain rule that $\Omega$ is $\scal_\lambda$-invariant, i.e., that $\w$ is $\Scal_\lambda$-invariant.

\subsection*{THE GNS REPRESENTATION}
We define the $^*$-representation $\rho\colon \cA\rightarrow \Borel(\Hil)$ in analogy to \eqref{pdisisad}, by linear extension of
\begin{align}
\label{odood}
\rho(o_1\ot{\dots}\ot o_n):= \wt{o}_1\cp{\dots}\cp \wt{o}_n.
\end{align} 
 By the same reasons as above, $\rho$ is zero on $\ID$; thus, defines a $^*$-representation $\varrho\colon \aA\rightarrow \Borel(\Hil)$. Obviously, $\chi_0\in \Hil$ is cyclic, and we have
\begin{align*}
	\w(a)=\langle\chi_0, \varrho(a)(\chi_0)\rangle\quad\forall\: a\in \aA
\end{align*}  
by construction. Thus, $\varrho$ is unitarily equivalent to the GNS-representation that corresponds to $\w$, just by what we have discussed in the last part of Sect.\ \ref{quanti}. 

Let us finally clarify the connection between the representations in the standard and the Fleishhack case. For this, let $\cA, \aA, \ID, \w, \rho, \varrho$ be defined as in the second case, and denote by $\cA_s, \aA_s, \ID_s,\w_s,\rho_s, \varrho_s$ the respective quantities in the standard one. 
It is then obvious from the definition of $\w$ that $\ID_0\subseteq \ID_\w$ holds, for $\ID_0\subseteq \aA$ the both-sided ideal generated by $\wh{\B}_0:=\{\wh{\vp}_0\:|\: \vp_0\in \B_0\}$ \footnote{Alternatively, this can also be obtained from the properties of a dilation-invariant state that we have derived in the second part of Sect.\ \ref{jndshjsd}.}. This means that in the GNS-representation of $\w$, the elements of $\ID_0$ correspond to zero operators, as well as to zero-Hilbert states; which is the reason why we could use the same pre-Hilbert space $\Hil$ for the definition of $\varrho$ and $\varrho_s$. 
More precisely, the canonical inclusion $\inn\colon \cA_s\rightarrow \cA$ carries over to an injection $\iota\colon \aA_s\rightarrow \aA$, and then we have
\begin{align*}
	\aA= \iota(\aA_s)\oplus\ID_0\quad\text{with}\quad 
	\varrho\cp \iota=\varrho_s\quad\text{and}\quad\varrho|_{\ID_0}=0,
\end{align*} 
cf.\ Appendix B. Thus, the GNS-representation in the Fleischhack case is actually just given by the standard one.

\section{EXTENSION: THE BIANCHI I CASE}

The key new elements in this paper, which have allowed all residual diffeomorphisms to have a well-defined action in the quantum 
cosmological model, are (1.) to let residual diffeomorphisms act not only on the dynamical variables, 
but also on the fiducial cell, and (2.) the use of the \textit{canonical} 
momenta conjugate to the connection variable in defining the quantum algebra.
Here we apply these same two elements to the case of Bianchi I quantum cosmology. 
We show that again one obtains
an action of \textit{all} residual diffeomorphisms on the quantum algebra, 
and not just an action of volume preserving ones.
Furthermore, we show how invariance, under this action, of a GNS state on the algebra, 
yields the unique state corresponding to the standard representation of Bianchi I defined in Ashtekar and Wilson-Ewing \cite{aw2009}.  
Thus, we demonstrate that the approach of this paper leads to
the same conclusions as those in Ashtekar and Campiglia \cite{UniquOfKinLQC},
whence they may be viewed as generalizing the results of \cite{UniquOfKinLQC}
in such a way that the isotropic case --- as well as the Fleischhack case --- can also be handled.

The Bianchi I cosmological model is obtained by starting with the same full theory framework reviewed in 
Sect.\ \ref{sec:classical}, but then imposing invariance under only the translation subgroup of the Euclidean group.
If one additionally uses diffeomorphism and gauge rotation freedom to impose that the triad be diagonal in the fixed coordinate system ---
$e^i_\mathbf{a} = v_i \delta^i_\mathbf{a}$, the so-called diagonal gauge --- then all terms in the Lagrangian involving 
time derivatives reduce to  
\begin{align}
\label{syltermBI}
&\sum_{i=1}^3 \frac{V_0[\mathcal{C}]}{k\gamma}\frac{|v_1 v_2 v_3|}{v_i} \frac{\mathrm{d}}{\mathrm{d}x^{\mathbf{0}}} 
\left( \gamma\he\omega_{\mathbf{i}}^{i0} - \frac{1}{2}\epsilon^{ijk} \omega_{\mathbf{i}jk}\right)
 \end{align}
where the unbold and bold indices $i$ and $\mathbf{i}$ have the same numerical value, 
the bold $\mathbf{i}$ denotes component with respect to $x^\mathbf{i}$,
and all fields are independent of the spatial variables $x^\mathbf{a}$.
Thus, 
\begin{align}
\label{BIcanvars}
c^i := \gamma\he\omega_\mathbf{i}^{i0}-\frac{1}{2} \epsilon^{ijk} \omega_{\mathbf{i}jk},
\qquad
p_i := \frac{V_0[\mathcal{C}]}{k \gamma}\frac{|v_1 v_2 v_3|}{v_i} 
\end{align}
are the dynamical variables and their canonical conjugates, with Poisson brackets 
$\{c^i, p_j\}= \delta^i_j$.
The only components of $\omega_\alpha^{IJ}$ appearing in
the above symplectic term (\ref{syltermBI}) are $c^i$, so that the rest of its components are Lagrange multipliers. 
Varying the action with respect to these Lagrange multipliers implies, in particular, the constraint that the Ashtekar connection
\begin{align*}
A^i_{\mathbf{a}} := -\left(\frac{1}{2}\epsilon_{ijk}\, \omega_\mathbf{a}^{jk} + \gamma \,  \omega_\mathbf{a}^{0i}\right)
\end{align*}
be diagonal, i.e., $A^i_\mathbf{a} = 0$ for $\mathbf{a} \neq i$.  
This constraint is second class, 
as we also trivially have the constraint on the conjugate momentum $\Pi_{A^i_{\mathbf{a}}} = 0$ for $\mathbf{a} \neq i$,
and $\{A^i_\mathbf{a}, \Pi_{A^i_{\mathbf{a}}} \} = 1$. Hence it must be solved prior to quantization, 
yielding an Ashtekar connection of the form
\begin{align*}
A^i_\mathbf{a} = c^i \delta^i_\mathbf{a} .
\end{align*}

The ``residual diffeomorphisms'' in this Bianchi I framework consist in the following three dimensional group of 
(possibly) \textit{anisotropic} dilations. 
For each $\vec{\lambda} \in (\mathbb{R}_{>0})^3$,
let $\Phi_{\vec{\lambda}}$ denote the diffeomorphism which expands the $x^{\mathbf{a}}$ direction by a factor of
$\lambda^{\mathbf{a}}$ for $\mathbf{a} = 1,2,3$, as measured in the fixed coordinate system $x^{\mathbf{a}}$.
Letting $\Phi_{\vec{\lambda}}$ act on the cell $\mathcal{C}$ maps
$V_0[\mathcal{C}]$ to $|\lambda_1 \lambda_2 \lambda_3| V_0[\mathcal{C}]$. 
Furthermore, $\Phi_{\vec{\lambda}}$ acts on the co-vectors
$e^i_{\mathbf{a}} = v^i \delta^i_{\mathbf{a}}$ and $A^i_{\mathbf{a}} = c^i \delta^i_\mathbf{a}$
via pull-back by $\Phi_{\vec{\lambda}}^{-1}$, inducing the action
$(v^i, c^i) \mapsto (\lambda_i^{-1} v^i, \lambda_i^{-1} c^i)$, which, via (\ref{BIcanvars}), translates into the action
\begin{align}
\label{canvaraction}
(p_i, c^i) \mapsto (\lambda_i \, p^i, \lambda_i^{-1} c^i)
\end{align}
of the dilations on the basic variables $(p_i, c^i)$.

In the Bianchi I approach of \cite{UniquOfKinLQC}, 
 holonomies of the Ashtekar connection $A^i_\mathbf{a} = c^i \delta^i_\mathbf{a}$ along straight curves parallel to one of the axes of the fixed coordinate system are considered in order to define the configuration algebra $\cC$, yielding $\CAPP$. 
This is generated by the functions $\chi_\lambda\cp\prr_i$ for $\lambda\in \RR$ and $i=1,2,3$,
 where $\prr_i$ denotes projection to the $i^{\text{th}}$ component in $\mathbb{R}^3$.  
 We let $\VDZ$ denote the span of these functions, in analogy to \eqref{uoioioi}. 
 
The classical Poisson algebra is then the complex vector space $\CC\cdot p_1\times \CC\cdot p_2\times\CC\cdot p_3 \times \DIDE$ 
with bracket
\begin{align}
\label{dsopspodfgfgf}
\{(z p_i,\vp),(z'p_j,\vp')\}=\{0,z'\partial_j\vp-z\partial_i\vp'\}
\end{align}
for $1\leq i,j\leq 3$: here, $\DIDE\subseteq \cC$ is chosen to be the largest subalgebra of $\cC$ for which \eqref{dsopspodfgfgf} makes sense; namely that consisting of all $\vp\in \cC\cap C^\infty(\RR^3)$ such that
\begin{align*}
	(\partial_{k_1}\cp\dots\cp \partial_{k_n})(\vp)\in \cC\quad\text{ for }\quad 1\leq k_1,\dots,k_n \leq 3
\end{align*}
for each $n\in \NN$.
This is dense in $\cC$ as clearly $\VDZ\subseteq \DIDE$ holds.
Already if one additionally takes holonomies along straight curves not-parallel to the axes of the fixed coordinate system into account, existence of such dense $\DIDE$ closed under partial derivatives is by far not clear. For this reason, and because the Fleischhack algebra has not been calculated for Bianchi I so far, we do not consider the analogue of the Fleischhack generalization for the Bianchi I case.

The quantum holonomy-flux $^*$-algebra $\aA$ is the quotient of the (mixed) complex tensor algebra $\cA$, 
over the vector spaces $\DIDE$ and $\CC\cdot p_i$ for $i=1,2,3$, by the both sided $^*$-ideal that is generated by 
\begin{align*}
	 p_j\ot\varphi-\varphi&\ot p_j+\I\cdot\partial_j\varphi\\[2pt]
	 \varphi_1\ot\varphi_2- \varphi_1\varphi_2&\quad\:\: p_i\ot p_j-p_j\ot p_i\\[2pt]
	 	\chi_0 &- 1_\cA \hspace{48pt}
\end{align*}
for $\varphi,\varphi_1,\varphi_2\in \DIDE$ and $i,j=1,2,3$. 
We let $\p_i:=[ p_i]$ for $i=1,2,3$, $\E:=[ \chi_0]$, and $\wh{\vp}:=[ \vp]$ for $\vp\in \DIDE$,
so that $[\p_i, \p_j] =0$ for $1\leq i,j\leq 3$ as well as
\begin{align*}
\textstyle[\p_i, \widehat{\varphi}] = - \I \cdot \widehat{\partial_i\varphi}\qquad\forall\:\varphi\in \DIDE.
\end{align*} 
The action of dilations (\ref{canvaraction}) induces an action $\scal_{\vec{\lambda}}: \cA \rightarrow \cA$
for each $\vec{\lambda} = (\lambda_1, \lambda_2, \lambda_3)$ via 
$\scal_{\vec{\lambda}}(z p_i):=\lambda_i\cdot z p_i$ for $z\in \CC$ and $i=1,2,3$, $\scal_{\vec{\lambda}}(1_\cA):=1_\cA$, as well as 
\begin{align*}  
	\scal_{\vec{\lambda}}(\varphi)\colon (x_1,x_2,x_3)\mapsto \vp( x_1/\lambda_1,x_2/\lambda_2,x_3/\lambda_3)
\end{align*}
for each $\vp\in \DIDE$. This action lifts to an action $\Scal_{\vec{\lambda}}\colon \aA\rightarrow\aA$ 
on the quantum algebra.
If $\w\colon \aA\rightarrow \cC$ is a $\Scal_{\vec{\lambda}}$-invariant state for all
$\lambda_1, \lambda_2, \lambda_3 > 0$, then the same arguments as in the homogeneous isotropic case show that it is uniquely determined by the continuous linear functional $\w_{\DIDEZ}\colon \DIDEZ\rightarrow \CC$,\: $\vp\mapsto \w(\wh{\vp})$ 
for $\DIDEZ:=\DIDE$ as we have $\wh{p}_i\in\ID_\w$ (the Gel'fand ideal) for $i=1,2,3$. 
Moreover, in manner similar to (\ref{gfhghg}),
we find that $\w_{\DIDEZ}(\E)=1$ as well as $\w_{\DIDEZ}(\chi_\lambda\cp\prr_i)=0$ holds for $\lambda\neq 0$ and $i=1,2,3$. Uniqueness is then clear from continuity of $\w_{\DIDEZ}$,
which follows from the same arguments as in Sect.\ \ref{jndshjsd}.
For existence, we can just substitute \eqref{hggffhjhjkj} by
\begin{align*}
	\textstyle L\colon \cC\rightarrow \CC,\quad \vp\mapsto \lim_n \frac{1}{(2n)^3}\int_{[-n,n]^3}\vp(x)\:\dd x,
\end{align*}
and define $\langle\cdot,\cdot\rangle$ as in \eqref{fddfgfsss} as well as $\Hil:=\DIDEZ=\DIDE$ just as in the homogeneous isotropic case \footnote{Again, we have $\langle\vp,\vp \rangle>0$ for each $0\neq\vp\in \CAPP$: this just follows as in Appendix A if we replace $\RB\equiv\Spec(\CAP)$ by $\RBB\equiv\Spec(\CAPP)$ in the argumentation there.}. 
We define the operators $\wt{z1}_\cA:=z\cdot \id_\Hil$ for each $z\in \CC$, as well as
\begin{align*}
	\wt{\vp}\colon \Hil\rightarrow\Hil,&\quad \vp'\mapsto \vp\cdot\vp'\\
	\wt{z p_j}\colon\Hil\rightarrow\Hil,&\quad \vp'\mapsto -z\I\cdot \partial_j\vp'
\end{align*}
for $j=1,2,3$, each $\vp\in \DIDE$ and $z\in \CC$. Then, $\Omega: \cA \rightarrow \mathbb{C}$ 
defined in exact analogy to \eqref{pdisisad} 
carries over to a state on $\aA$ that has the desired properties.
The corresponding GNS-representation then is unitarily equivalent to that defined  by expression \eqref{odood}, 
which one sees to be the standard representation of Bianchi I
introduced by Ashtekar and Wilson-Ewing \cite{aw2009}.

\section*{ACKNOWLEDGMENTS}
We thank an anonymous referee for suggesting that we include an analysis of the Bianchi I case. We furthermore thank Christian Fleischhack for communicating to us \cite{CSL2} the technical oversight we have made in the continuity part in Sect.\ \ref{jndshjsd} in the  earlier versions of this article. 
This work has been supported in part by the Alexander von Humboldt foundation of Germany and NSF Grant PHY-1505490.

\section*{APPENDIX A}
Let us first observe that 
\begin{align*}
	\textstyle L(\vp)= \int \Gel(\vp)\:\dd\muRB \quad\forall\:\vp\in \CAP
\end{align*}
holds for $\muRB$ the Haar measure on the Bohr compactification $\RB\equiv\Spec(\CAP)$ of $\RR$, and
\begin{align*}
	 \Gel\colon \CAP&\rightarrow  C(\RB)\\
	 	\vp&\mapsto [\wh{\vp}\colon \alpha\mapsto \alpha(\vp)]
\end{align*} 
 the corresponding Gel'fand transform. This  
 is immediate for each $\vp\in \VDZ$ from general theory of Haar measures on compact abelian groups; and thus, by continuity, clear for all $\vp\in \CAP$. 
 Consequently, we have 
 \begin{align*}
 \textstyle 0=\langle \vp,\vp\rangle=\int |\wh{\vp}|^2\:\dd\muRB\quad&\Longrightarrow\quad \wh{\vp}=0\\
 &\Longrightarrow\quad \vp=0.
 \end{align*}
 This is because $\wh{\vp}$ is continuous, and because $\muRB(O)>0$ holds for each non-empty open subset $O\subseteq \RB$. 

\section*{APPENDIX B}
First observe that $\inn\colon \cA_s\rightarrow \cA$ carries over to a map $\iota\colon \aA_s\rightarrow \aA$ because $\inn(\ID_s)\subseteq \ID$ holds, so that 
\begin{align*}
	\rho\cp \inn=\rho_s\quad\Longrightarrow\quad \varrho\cp \iota=\varrho_s.
\end{align*}
Then, to verify injectivity of $\iota$, we define $\pr_s\colon \cA\rightarrow \cA_s$ by linear extension of
\begin{align*}
	\pr_s(o_1\ot{\dots}\ot o_n):= \pr(o_1)\cp{\dots}\cp \pr(o_n)
\end{align*} 
with $\pr(\vp):=\pap(\vp)$ for each $\vp\in \DIDE$, and
\begin{align*}
	 \pr(z p):=zp\hspace{1pt},\quad \pr(z1_\cA):=z1_\cA\qquad\forall\: z\in \CC.
\end{align*}
Since $\pr_s(\ID)=\ID_s$ holds, $\pr_s$ carries over to a map $\pi\colon \aA\rightarrow \aA_s$; and then
\begin{align*}
	\pr_s\cp\inn=\id_{\cA_s}\quad\Longrightarrow\quad\pi\cp\iota=\id_{\aA_s}
\end{align*}
shows that $\iota$ is injective, and that  
$\inn\cp \pr_s$ 
is a projection. We thus have 
\begin{align*}
	\cA=(\inn\cp \pr_s)(\cA)\oplus \ker[\inn\cp \pr_s]=\inn(\cA_s)\oplus \ker[\pr_s],	
\end{align*}
whereby $\ker[\pr_s]$ is easily seen to be the both-sided ideal $J_0\subseteq \cA$ generated by $\B_0$. Then, applying the quotient map $\quot\colon \cA\rightarrow \aA$ to both sides, we find that
\begin{align*}
	\aA= (\quot\cp \inn)(\cA_s)\oplus \quot(J_0)=\iota(\aA_s)\oplus \ID_0
\end{align*}
holds. Here, in the first step, the direct sum property is preserved, because
\begin{align*}
	a\in (\quot\cp \inn)(\cA_s)\cap \quot(J_0)\quad&\Longrightarrow\quad \quot(\inn(a_s))=a=\quot(a_0)\\
	&\Longrightarrow\quad \inn(a_s)-a_0\in \ID\\
	&\Longrightarrow\quad a_s=(\pr_s\cp\inn)(a_s)\in \ID_s\\
	&\Longrightarrow\quad \inn(a_s)\in \ID\\
	&\Longrightarrow\quad a=0,
\end{align*}
for some $a_s\in \cA_s$ and $a_0\in J_0=\ker[\pr_s]$.

\end{document}